\documentclass[12pt,twoside]{article}   %For printing on two sides of page
\usepackage[super,sort,comma]{natbib}
\usepackage{fancyhdr}		%Gives headers and footers defined below
\usepackage[section]{placeins}   %
\usepackage{graphicx}

\makeatletter \renewcommand\@biblabel[1]{$^{#1}$} \makeatother
\newcommand{\cen}[1]{\begin{center} #1 \end{center}}

\usepackage[mathlines]{lineno}
\usepackage{hyperref}
\hypersetup{ colorlinks,
    citecolor=blue,
    filecolor=blue,
    linkcolor=blue,
    urlcolor=blue
}

\usepackage{xcolor}
\definecolor{gray}{rgb}{0.6,0.6,0.6}
\definecolor{red}{rgb}{0.85,0,0}
\definecolor{green}{rgb}{0,0.85,0}
\definecolor{blue}{rgb}{0,0,0.85}
\definecolor{beige}{rgb}{0.92,0.87,0.78}
\usepackage[all]{hypcap}    %causes link to figures to go to figure, not caption
\usepackage{amsmath,amssymb,amsfonts}% own package

\begin{document}

\cen{\sf {\Large {\bfseries Ultra Low-Parameter Denoising: Trainable Bilateral Filter Layers in Computed Tomography } \\  
\vspace*{10mm}
Fabian Wagner$^{1}$, Mareike Thies$^{1}$, Mingxuan Gu$^{1}$, Yixing Huang$^{1}$, Sabrina Pechmann$^{2}$, Mayank Patwari$^{1}$, Stefan Ploner$^{1}$, Oliver Aust$^{3}$, Stefan Uderhardt$^{4, 5}$, Georg Schett$^{4, 5}$, Silke Christiansen$^{2, 6}$, Andreas Maier$^{1}$} \\
$^{1}$Pattern Recognition Lab, Friedrich-Alexander University Erlangen-Nürnberg, Germany\\
$^{2}$Fraunhofer Institute for Ceramic Technologies and Systems IKTS, Forchheim, Germany\\
$^{3}$Leibniz Institute for Analytical Sciences ISAS, Dortmund, Germany\\
$^{4}$Department of Internal Medicine 3 - Rheumatology and Immunology, Friedrich-Alexander University Erlangen-Nürnberg, Germany\\
$^{5}$University Hospital Erlangen, Germany\\
$^{6}$Institute for Nanotechnology and Correlative Microscopy e.V. INAM, Forchheim, Germany
\vspace{5mm}\\
%Version typeset \today\\
}

\pagenumbering{roman}
\setcounter{page}{1}
\pagestyle{plain}
Author to whom correspondence should be addressed. email: fabian.wagner@fau.de\\
% note, probably best not to use a student's e-mail as it won't be valid for
% very long.

\begin{abstract}
\noindent {\bf Background:} Computed tomography is widely used as an imaging tool to visualize three-dimensional structures with expressive bone-soft tissue contrast. However, CT resolution and radiation dose are tightly entangled, highlighting the importance of low-dose CT combined with sophisticated denoising algorithms.\\
{\bf Purpose:} Most data-driven denoising techniques are based on deep neural networks and, therefore, contain hundreds of thousands of trainable parameters, making them incomprehensible and prone to prediction failures. Developing understandable and robust denoising algorithms achieving state-of-the-art performance helps to minimize radiation dose while maintaining data integrity.\\
{\bf Methods:} This work presents an open-source CT denoising framework based on the idea of bilateral filtering. We propose a bilateral filter that can be incorporated into a deep learning pipeline and optimized in a purely data-driven way by calculating the gradient flow toward its hyperparameters and its input. Denoising in pure image-to-image pipelines and across different domains such as raw detector data and reconstructed volume, using a differentiable backprojection layer, is demonstrated.\\
{\bf Results:}Although only using three spatial parameters and one range parameter per filter layer, the proposed denoising pipelines can compete with deep state-of-the-art denoising architectures with several hundred thousand parameters. Competitive denoising performance is achieved on x-ray microscope bone data (0.7053 and 33.10) and the 2016 Low Dose CT Grand Challenge dataset (0.9674 and 43.07) in terms of SSIM and PSNR. \\
{\bf Conclusions:} Due to the extremely low number of trainable parameters with well-defined effect, prediction reliance and data integrity is guaranteed at any time in the proposed pipelines, in contrast to most other deep learning-based denoising architectures.\\

\end{abstract}

%\newpage     %may or may not be needed

%The table of contents is for drafting and refereeing purposes only. Note
%that all links to references, tables and figures can be clicked on and
%returned to calling point using cmd[ on a Mac using Preview or some
%equivalent on PCs (see View - go to on whatever reader).
%\tableofcontents

\newpage

\setlength{\baselineskip}{0.7cm}      %double spacing		

\pagenumbering{arabic}
\setcounter{page}{1}
\pagestyle{fancy}
\section{Introduction}
\label{sec:introduction}

Ionizing radiation used in Computed Tomography (CT) is known to be harmful to any living tissue. Therefore, the amount of deposited energy in each investigated sample, the so-called dose, is to be minimized following the \textit{As Low As Reasonably Achievable} (ALARA) principle \cite{boone2012radiation}. However, radiation dose and image resolution are inherently entangled as a certain number of x-rays must be absorbed by the scanned patient to achieve the desired contrast \cite{barrett1976statistical}. Further reducing the patient dose results in degraded image quality due to increased Poisson noise in CT projections.\\
Since the emergence of the first CT scanners, denoising algorithms were developed and applied to restore image quality while keeping the radiation dose moderate \cite{maier2015gpu}. Non-linear filters \cite{kelm2009optimizing, maier2011three, manduca2009projection, manhart2014guided} and iterative reconstruction techniques \cite{han2010algorithm, beister2012iterative, gilbert1972iterative} have been successfully applied. Whereas such approaches usually require hand-tuned hyperparameters, cannot abstract complex features, or are known to be computationally expensive, purely data-driven, end-to-end trainable, approaches have been proposed in recent years \cite{chen2017low, patwari2020, fan2019quadratic, yang2018low, kang2018deep, kang2017deep, ketcha2021} fueled by the emergence of deep learning and, in particular, convolutional neural networks. Most of these models achieve competitive denoising performance but are built on deep neural networks with multiple layers and contain hundreds of thousands of trainable parameters. Although deep architectures help networks extract complex features from data, such models are often regarded as black boxes as it is impossible to fully comprehend their data processing and control failing network predictions. Besides, adversarial examples in the form of small perturbations of the network input can lead to undesired drastic changes in the prediction due to the high dimension of extracted features \cite{yuan2019adversarial}. Such uncertainties often prohibit deep learning applications in the medical imaging field where the reliability of the data must be maintained \cite{antun2020instabilities, huang2018some}. Additionally, training large numbers of parameters in neural networks usually requires broad medical datasets with paired ground truth data, which can be hard to obtain.\\
The bilateral filter has been successfully applied on CT data \cite{manduca2009projection} as it performs combined filtering in spatial and range domain with Gaussian kernels smoothing the noise fluctuations while preserving edges \cite{tomasi1998bilateral}. However, hyperparameters---fundamentally determining the filter performance---usually have to be hand-picked. Multiple works have been proposed aiming for automatically finding optimal filter parameters \cite{chen2013optimization, anoop2019medical, kishan2012optimal, dai2018content, peng2010bilateral} using risk estimators or grasshopper optimization. All these techniques are not suitable for integration into an end-to-end, data-driven, optimization pipeline and are based on different statistical assumptions on the noise. A different work presents a bilateral filter based on convolutional filtering the permutohedral lattice, a higher-dimensional representation of the image data, that can be optimized \cite{jampani2016learning}. However, their algorithm requires so-called \textit{splatting} and \textit{slicing} operations, which can only approximate the data and, therefore, introduce uncertainties. Additionally, the requirement of a higher-dimensional grid increases computational demands. Patwari \textit{et al.} \cite{patwari2020} proposed the JBFnet, a deep neural network architecture inspired by joint bilateral filtering that can be trained in a purely data-driven way. They showed competitive performance to much deeper networks, although reducing the number of trainable parameters by choosing a shallow convolutional architecture. Besides, denoising approaches indirectly optimizing hyperparameters of (joint) bilateral filters using reinforcement learning have been proposed \cite{kang2021low, patwari2020low}. However, their training is more sophisticated as it is again based on deep architectures for choosing the correct parameter updates. Additionally, Patwari \textit{et al.} \cite{patwari2020low} make use of a residual network to generate an image prior as well as a reward network.\\
We aim to extend the aforementioned approaches by proposing a trainable bilateral filter layer using only the inherent four filter parameters (including three spatial filter dimensions) during training and inference. By analytically deriving the gradient flow toward the parameters as well as to the layer input, we can directly optimize all filter parameters via backpropagation and incorporate the C++/CUDA accelerated layer into any trainable pipeline using the \textit{PyTorch} framework \cite{pytorch2019}. Additionally, we show simultaneous optimization of filter parameters in the projection and image domain, using a differentiable backprojection layer \cite{syben2019}. Due to the very low number of trainable parameters, we can optimize our pipeline with only very little training data and in a self-supervised manner using Noise2Void training \cite{krull2019noise2void}, while still achieving competitive performance compared to state-of-the-art deep neural networks. We explain the competitive denoising performance with the theoretical findings of Maier \textit{et al.} \cite{maier2019, maier2018precision}, who proved that incorporating prior knowledge into artificial neural networks lowers the upper error bound of the model prediction. The experiments are performed on the 2016 Low Dose CT Grand Challenge dataset \cite{mccollough2017low} ($25\,\%$ dose) for benchmark purposes as well as on a low-dose x-ray microscope (XRM) bone dataset ($10\,\%$ dose).

\section{Materials and Methods}

\subsection{Trainable bilateral filter}
The bilateral filter had great success in CT denoising due to its ability to smooth image content while preserving edges. This is achieved by a composed filter kernel with spatial and range contributions. With the noisy input reconstruction $\mathbf{X}$ and the denoised prediction $\mathbf{\hat{Y}}$, the discrete filter operation can be written as \cite{tomasi1998bilateral}
\begin{align}
    \hat{Y}_k = \frac{1}{w_k} \underbrace{\sum_{n \in \mathcal{N}} G_{\sigma_s}(\mathbf{p_k} - \mathbf{p_n}) G_{\sigma_r}(X_k - X_n) X_n}_{\substack{=:\,\alpha_k}}
    \label{eq:DefBF}
\end{align}
with the definition of the normalization factor $w_k$
\begin{align}
    w_k := \sum_{n \in \mathcal{N}} G_{\sigma_s}(\mathbf{p_k} - \mathbf{p_n}) G_{\sigma_r}(X_k - X_n),
\end{align}
voxel index $k \in \mathbb{N}$, and the Gaussian function
\begin{align}
    G_{\sigma_r}(c) := \exp \left(-\frac{c^2}{2\sigma_r^2}\right).
\end{align}
Each predicted output voxel $\hat{Y}_k$ is calculated from the neighborhood $\mathcal{N}$ of voxel $X_k$, indexed by $n \in \mathcal{N}$. The spatial kernel $G_{\sigma_s}$ performs image smoothing and is dependent on the distance between the positions $\mathbf{p_k} \in \mathbb{N}^d$ and $\mathbf{p_n} \in \mathbb{N}^d$. In the $d$-dimensional case $G_{\sigma_s}$ is composed of multiple Gaussian kernels, e.g.,
\begin{align}
    G_{\sigma_s}(\mathbf{c}) = \prod_{\beta \in \{x,y,z\}} \exp \left(-\frac{c_\beta^2}{2\sigma_\beta^2}\right)
\end{align}
assuming $d=3$ spatial filter dimensions and the spatial distance vector $\mathbf{c} = (c_x, c_y, c_z)$. The additional range kernel $G_{\sigma_r}$ is responsible for preserving edges during filtering as similar voxel values are stronger weighted by incorporating the intensity distance $(X_k - X_n)$.\\

\begin{figure}[ht]
\centerline{\includegraphics[trim=0.2cm 0.6cm 0.5cm 0cm, width=0.75\columnwidth]{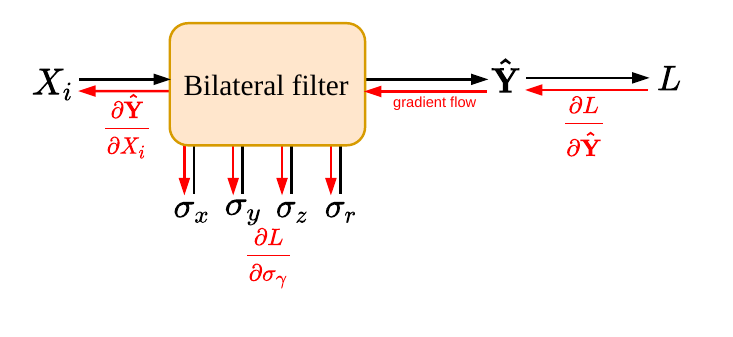}}
    \caption{Working scheme of the proposed trainable bilateral filter layer. Black arrows mark the forward pass, whereas red arrows illustrate the gradient flow toward the input $X_i$ and the filter parameters $\sigma_\gamma$ during backpropagation.}
    \label{fig:BF_gradient_flow}
\end{figure}

Kernel widths $\sigma_s$ and $\sigma_r$ are hyperparameters of the bilateral filter and are usually hand-picked by the user dependent on image content and voxel range. However, tuning the filter layers by hand is cumbersome, and finding an optimal parameter set can not be guaranteed. To optimize the parameters in a data-driven fashion and incorporate them in a fully differentiable pipeline, the gradient, given by the derivative of the loss function $L$ with respect to each parameter $\sigma_\gamma$ must be calculated
\begin{align}
    \frac{\partial L}{\partial \sigma_\gamma} = \frac{\partial L}{\partial \mathbf{\hat{Y}}} \frac{\partial \mathbf{\hat{Y}}}{\partial \sigma_\gamma}  = \sum_k \frac{\partial L}{\partial \hat{Y}_k} \frac{\partial \hat{Y}_k}{\partial \sigma_\gamma}.
\end{align}
Automatically deriving the gradient with a framework like \textit{PyTorch} is in principle possible, but comes with a huge computational and memory overhead. It would require calculating gradients for each computation in the filter forward pass, which is computationally infeasible with conventional hardware in a reasonable run time. Instead, we directly calculate the analytical solution of $\frac{\partial \hat{Y}_k}{\partial \sigma_\gamma}$, which is described in detail in the Appendix. Additionally, the gradient with respect to every voxel of the noisy input reconstruction $X_i$ needs to be derived to propagate a loss into previous filter layers during backpropagation to allow stacking multiple bilateral filters or incorporating the layer into a deep architecture
\begin{align}
    \frac{\partial L}{\partial X_i} = \frac{\partial L}{\partial \mathbf{\hat{Y}}} \frac{\partial \mathbf{\hat{Y}}}{\partial X_i} = \sum_k \frac{\partial L}{\partial \hat{Y}_k} \frac{\partial \hat{Y}_k}{\partial X_i}.
\end{align}
With each predicted $\hat{Y}_k$ being dependent on intensity differences of two input voxels $X_k$ and $X_n$, the analytical calculation of the gradient flow toward the filter input
\begin{align}
    \frac{\partial \hat{Y}_k}{\partial X_i} = - w_k^{-2} \alpha_k \frac{\partial w_k}{\partial X_i} + w_k^{-1} \frac{\partial \alpha_k}{\partial X_i}
\end{align}
is more complex than the derivative with respect to the kernel widths $\sigma_\gamma$. Eventually, it requires distinguishing the two cases $k = i$ and $k \neq i$ for finding an applicable solution. They refer to the off-center and center elements of the bilateral filter kernel respectively. The detailed calculation is again conducted in the Appendix, accompanied by implementation notes. Fig. \ref{fig:BF_gradient_flow} illustrates the data flow in a bilateral filter layer. The forward and backward pass of the layer is implemented in C++ and CUDA to leverage performance and integrated into the \textit{PyTorch} framework using the \textit{pybind11} module \cite{pybind11}. Our open-source filter layer is publicly available under \url{https://github.com/faebstn96/trainable-bilateral-filter-source}.

\subsection{Denoising pipelines}

\begin{figure*}[tbh]
    \centering
    \includegraphics[width=\linewidth]{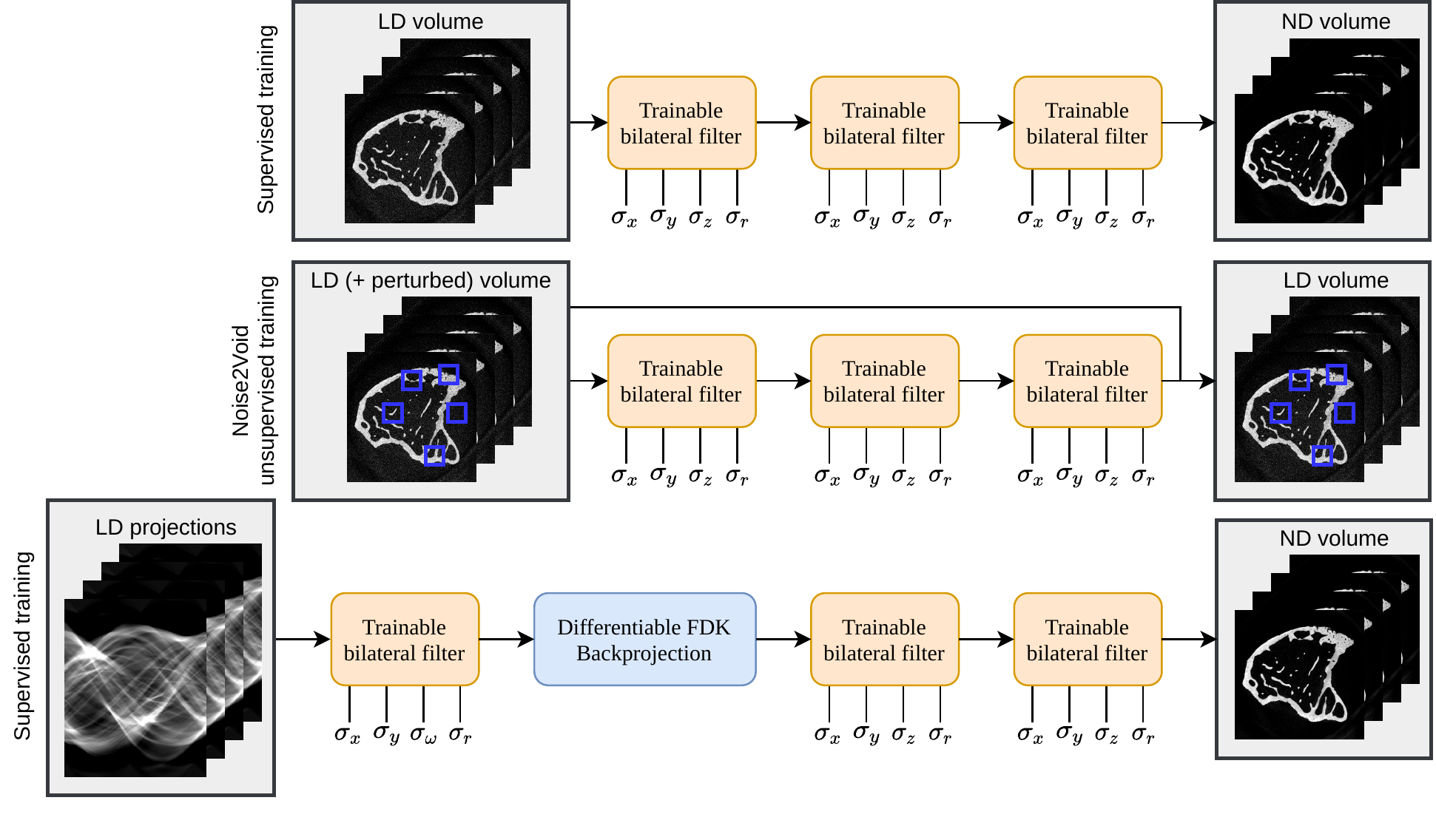}
    \caption{Illustration of the employed denoising pipelines containing the proposed bilateral filter layers. Multiple filters are trained in a supervised (first row) and unsupervised (Noise2Void \cite{krull2019noise2void}, second row) fashion. Additionally, bilateral filter layers are optimized in projection and image domain simultaneously using a differentiable backprojection operator (dual-domain BFs, third row). The $\sigma_\gamma$ parameters represent the only trainable parameters of the networks and are all optimized independently. ND refers to the normal-dose reconstructions whereas LD denotes the low-dose data incorporating noise.}
    \label{fig:BF_pipelines}
\end{figure*}

Multiple denoising pipelines containing trainable bilateral filter layers, illustrated in Fig. \ref{fig:BF_pipelines}, are employed to investigate the performance and limitations of the layers. First, a simple post-processing approach is studied, directly applying three subsequent bilateral filter layers on the reconstructed volume. The filters are optimized for the mean squared error (MSE) loss calculated between prediction $\mathbf{\hat{Y}}$ and normal-dose volume $\mathbf{Y}$.\\
Additionally, self-supervised training is performed using the Noise2Void (N2V) training scheme. The noisy volume is perturbed by randomly replacing a defined ratio of voxels ($1\,\%$) with voxel intensities from their respective $5^3$ voxel neighborhoods and fed through the denoising pipeline. Subsequently, an MSE loss is calculated between prediction and noisy non-perturbed volume at the voxel positions of the perturbations. If the noise contribution of neighboring voxels is approximately uncorrelated, the resulting gradient will converge to the denoised version of the volume, as proved by Krull \textit{et al.} \cite{krull2019noise2void}. The advantage of the N2V training scheme is that no paired high-dose data is required. Accordingly, no additional ground truth knowledge can be employed during training.\\
Finally, an FDK-based reconstruction pipeline with a single bilateral filter layer in the projection and two filters in the image domain is trained end-to-end on XRM data. The employed pipeline is illustrated in the last row of Fig. \ref{fig:BF_pipelines} and makes use of the fully differentiable reconstruction pipeline of Thies \textit{et al.} \cite{thies2021differentiable}, which is described in the following section. All filter parameters are optimized for the MSE loss between $\mathbf{\hat{Y}}$ and the normal-dose reconstruction $\mathbf{Y}$ in the image domain.

\subsection{XRM reconstruction pipeline}
The FDK-based \cite{feldkamp1984practical} reconstruction pipeline by Thies \textit{et al.} \cite{thies2021differentiable} connects acquired CT projection data with its 3D interpretable reconstruction. The backprojection is incorporated as a known operator\cite{maier2019}, using the differentiable layer provided by the \textit{Pyro-NN} framework by Syben \textit{et al.} \cite{syben2019}. The reconstruction pipeline is adapted to XRM projection data from a Zeiss Xradia 620 Versa microscope, which is a high-resolution cone-beam CT for small samples, and allows to propagate a loss calculated in the image domain to any location within the reconstruction pipeline enabled by its differentiable implementation. Accordingly, bilateral filter layers can be trained at multiple locations in the pipeline in a purely data-driven manner using the \textit{PyTorch} framework. Particularly, we perform experiments demonstrating the superior performance of bilateral filter layers applied in both projection and image domain simultaneously over denoising in image domain only.

\subsection{Experimental setup}
We compare the performance of the trainable bilateral filter layer-based pipelines to four deep state-of-the-art denoising architectures on the ten abdomen scans from the 2016 Low Dose CT Grand Challenge dataset \cite{mccollough2017low} with $1\,\text{mm}$ slice thickness reconstructed in slices of size $512 \times 512$. Here, patient \textit{L291} is selected for validation, patients \textit{L310, L333} and \textit{L506} for testing and all $3411$ slices from the remaining seven scans for training.\\
In a second experiment, we investigate the denoising performance on a dataset of high-resolution ex vivo XRM scans of mouse tibia bones. In the context of preclinical osteoporosis research, lacunae, bone structures with a diameter of 3-20$\,\text{\textmu m}$, were investigated \cite{hannah2010}. This is instructive for understanding the bone metabolism \cite{gruneboom20191, gruneboom20192} as they contain osteocytes, cells being heavily involved in the bone remodeling process \cite{gruber2008introduction, buenzli2015}. However, the severe amount of radiation dose required for resolving micrometer-sized lacunae in a mouse so far prohibits in vivo imaging \cite{mill2019towards, huang2021semi} and, therefore, requires sophisticated dose reduction techniques to push the XRM in vivo limit to the micrometer scale. We create a low-dose XRM dataset from ex-vivo mouse tibia bone samples to investigate in-vivo XRM acquisitions. The sample preparation was carried out as approved by the local animal ethics committee in compliance with all ethical regulations (license TS-10/2017). $1400$ high-resolution projections are acquired for every sample in a short-scan setting with $28\,\text{s}$ exposure time per projection image. Low-dose and normal-dose projections are reconstructed in a $2048^3$ grid of $1.4\,\text{\textmu m}$ isotropic voxel size using the pipeline of Thies \textit{et al.} \cite{thies2021differentiable}. Data acquired with $10\,\%$ dose is simulated by incorporating noise following Yu \textit{et al.} \cite{yu2012development}. Two separate stacks of $30$ slices each are reconstructed from every bone scan as the total amount of reconstructed slices would by far exceed the capacity of the denoising models and the variety of learnable local features. The final XRM bone dataset contains five scans for training, one for validation, and four for testing, which is in total equivalent to $9600$ $512 \times 512$ image patches, exceeding the size of the 2016 Low Dose CT Grand Challenge dataset.\\
The quadratic autoencoder architecture (QAE) proposed by Fan \textit{et al.} \cite{fan2019quadratic}, as well as the JBFnet by Patwari \textit{et al.} \cite{patwari2020}, both achieve remarkable results on the CT denoising task, outperforming many architectures proposed during the 2016 Low Dose CT Grand Challenge. The RED-CNN network, published by Chen \textit{et al.} \cite{chen2017low} is a deep, convolutional architecture with more than $1.8\,\text{Mio}$ parameters achieving state-of-the-art results likewise. Another denoising approach based on a generative adversarial network is WGAN \cite{yang2018low}, using a combination of Wasserstein and perceptual loss. Encouraged by their performance, these four methods are selected for comparison with multiple pipelines containing the proposed trainable bilateral filter layers on the aforementioned datasets.\\
In order to allow a fair comparison between the different denoising approaches, the networks are implemented and trained based on officially published code repositories and the description from their publications. The Adam optimizer with learning rates $4\cdot 10^{-4}$ (decaying), $1\cdot 10^{-4}$ (constant), $1\cdot 10^{-4}$ (decaying), and $1\cdot 10^{-5}$ (constant) is used for QAE, JBFnet, RED-CNN, and WGAN respectively. Both QAE and RED-CNN are trained using the MSE loss, whereas JBFnet makes use of a custom loss described in their publication \cite{patwari2020} as well as pre-training of its prior module. Generator and discriminator of WGAN are trained alternately using Wasserstein loss for stable convergence and a perceptual loss as presented in \cite{yang2018low}. All models are trained until the validation loss plateaus or increases.\\
In all bilateral filter-based models, $\sigma_{x,y,z}$ and $\sigma_{r}$ are initialized with $0.5$ and $0.01$ and optimized with two separate Adam optimizers with learning rates $0.01$ and $0.005$ respectively, as the spatial and range parameters operate on two independent scales. The tiny amount of required training data is demonstrated by training the bilateral filter pipelines only on a single stack of $21$ neighboring slices with a size of $512 \times 512$ voxels from one training scan of the Grand Challenge dataset. For the XRM bone dataset a stack of $15$ neighboring patches of size $512 \times 512$ voxels is used. Note that the bilateral filter layers cannot overfit the data assuming a comparable amount of noise within all scans due to the low number of trainable parameters with well-defined influence. Optimization is performed until convergence of the training loss for up to $5000$ iterations which took up to $20$ minutes (on an NVIDIA Quadro RTX 4000).\\

\begin{figure}[h]
\centerline{\includegraphics[width=0.5\columnwidth]{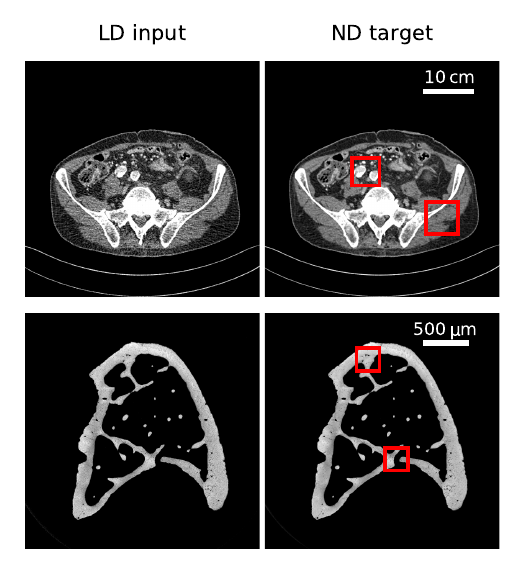}}
   \caption{Example input and target slices from the investigated Grand Challenge dataset ($25\,\%$ dose) and the XRM bone dataset ($10\,\%$ dose). Red squares highlight the ROIs for the visual comparison. The display windows are $[-150, 250]\,\text{HU}$ and $[0.25, 0.7]\,\text{arb.}\,\text{unit}$.}
   \label{fig:Overview_ROIs}
   \vspace*{2ex}
\end{figure}

\section{Results}

\subsection{Denoising results}

\begin{figure*}[htb]
    \centering
    \includegraphics[width=\linewidth]{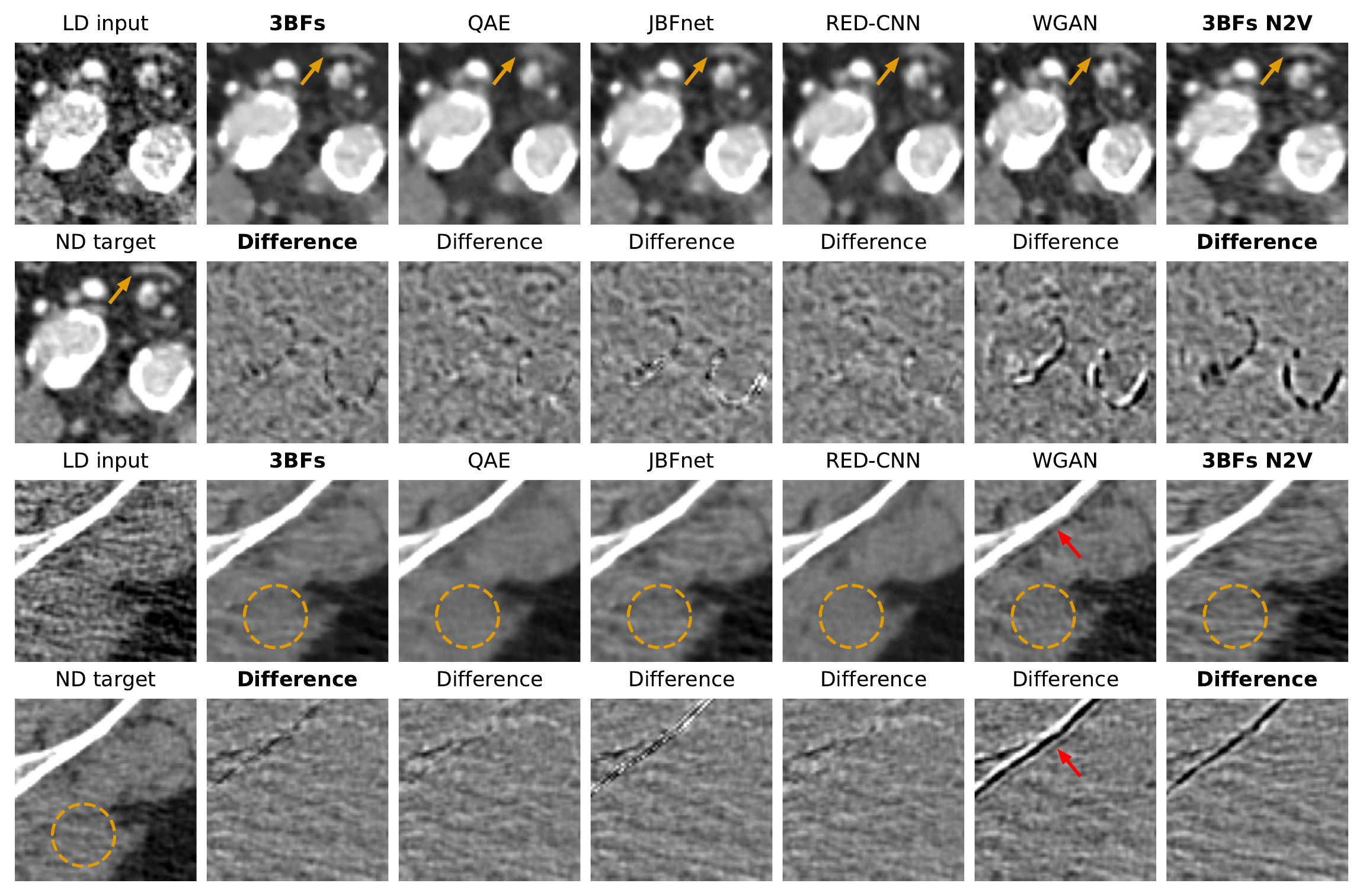}
    \caption{The first and third rows show denoising predictions from the 2016 Low Dose CT Grand Challenge dataset in the ROIs illustrated in Fig. \ref{fig:Overview_ROIs} for all employed methods. Our proposed pipelines are highlighted in bold letters. The display window is $[-150, 250]\,\text{HU}$. The second and fourth rows present difference images between prediction and target where prediction artifacts like blurred edges become particularly visible. Difference images are plotted in the window $[-150, 150]\,\text{HU}$.}
    \label{fig:results_rois_mayo}
    \vspace*{2ex}
\end{figure*}

\begin{table}[htbp]
\begin{center}
\caption{Quantitative denoising results of the compared pipelines on the test patients of the 2016 Low Dose CT Grand Challenge dataset. For each model, the number of trainable parameters and the average inference time per $512 \times 512$ image slice (Quadro RTX 4000) are presented. The names of our proposed pipelines and the best-performing models are highlighted.
\label{tab:quant_results_GC}
\vspace*{2ex}
}
\begin{tabular} {lccrc}
\hline
& \multicolumn{2}{c}{\textbf{Grand Challenge Dataset}} & & $[\text{s}]$\\
\cline{2-3}
& \textbf{SSIM} (mean $\pm$ std)&  \textbf{PSNR} (mean $\pm$ std) & \textbf{\# params} & \textbf{runtime}\\
\hline
Low-dose CT
& $0.8991 \pm 0.033$ & $38.64 \pm 1.62$ &  -  & -\\
\hline
WGAN \cite{yang2018low}
& $0.9341 \pm 0.085$ & $40.47 \pm 0.86$ & $7,800,000$ & \boldmath$0.1$ \\
RED-CNN \cite{chen2017low}
& \boldmath$0.9680 \pm 0.010$ & \boldmath$43.63 \pm 1.19$ & $1,800,000$ & $4.2$ \\
JBFnet \cite{patwari2020}
& \boldmath$0.9666 \pm 0.012$ & $42.77 \pm 1.13$ & $118,000$ & $5.6$ \\
QAE \cite{fan2019quadratic}
& $0.9650 \pm 0.010$ & \boldmath$43.35 \pm 1.18$ & $137,000$ & $2.0$ \\
\textbf{1 BF layer}
& $0.9656 \pm 0.013$ & $42.73 \pm 1.38$ & \boldmath$4$ & \boldmath$0.2$ \\
\textbf{3 BF layers}
& \boldmath$0.9674 \pm 0.012$ & \boldmath$43.07 \pm 1.42$ & \boldmath$12$ & \boldmath$0.5$ \\
\textbf{3 BFs (N2V \cite{krull2019noise2void})}
& $0.9577 \pm 0.015$ & $41.15 \pm 1.17$ & \boldmath$12$ & \boldmath$0.5$ \\
\hline
\end{tabular}
\end{center}
\end{table}

\begin{table}[hbtp]
\begin{center}
\caption{Quantitative denoising results of the compared pipelines on the test patients of the XRM bone dataset. For each model, the number of trainable parameters is shown in the right column.
\label{tab:quant_results_xrm}
\vspace*{2ex}
}
\begin{tabular} {lccr}
\hline
& \multicolumn{2}{c}{\textbf{XRM Bone Dataset}} & \\
\cline{2-3}
& \textbf{SSIM} (mean $\pm$ std)&  \textbf{PSNR} (mean $\pm$ std)& \textbf{\# parameters}\\
\hline
Low-dose CT
& $0.1652 \pm 0.0084$ & $19.95 \pm 0.27$ &  -  \\
\hline
WGAN \cite{yang2018low}
& $0.5170 \pm 0.0135$ & $29.45 \pm 0.23$ & $7,800,000$  \\
RED-CNN \cite{chen2017low}
& \boldmath$0.7122 \pm 0.0087$ & \boldmath$33.09 \pm 0.20$ & $1,800,000$  \\
JBFnet \cite{patwari2020}
& $0.6564 \pm 0.0139$ & $30.98 \pm 0.31$ & $118,000$  \\
QAE \cite{fan2019quadratic}
& \boldmath$0.7126 \pm 0.0088$ & \boldmath$33.09 \pm 0.20$ & $137,000$  \\
\textbf{1 BF layer}
& $0.6599 \pm 0.0167$ & $31.40 \pm 0.30$ & \boldmath$4$  \\
\textbf{3 BF layers}
& $0.6698 \pm 0.0158$ & $32.19 \pm 0.26$ & \boldmath$12$  \\
\textbf{3 BF layers (N2V \cite{krull2019noise2void})}
& $0.4590 \pm 0.0280$ & $27.61 \pm 0.57$ & \boldmath$12$  \\
\textbf{Dual-domain BFs}
& \boldmath$0.7053 \pm 0.0120$ & \boldmath$33.10 \pm 0.22$ & \boldmath$12$  \\
\hline
\end{tabular}
\end{center}
\end{table}

\begin{figure*}[htb]
    \centering
    \includegraphics[width=\linewidth]{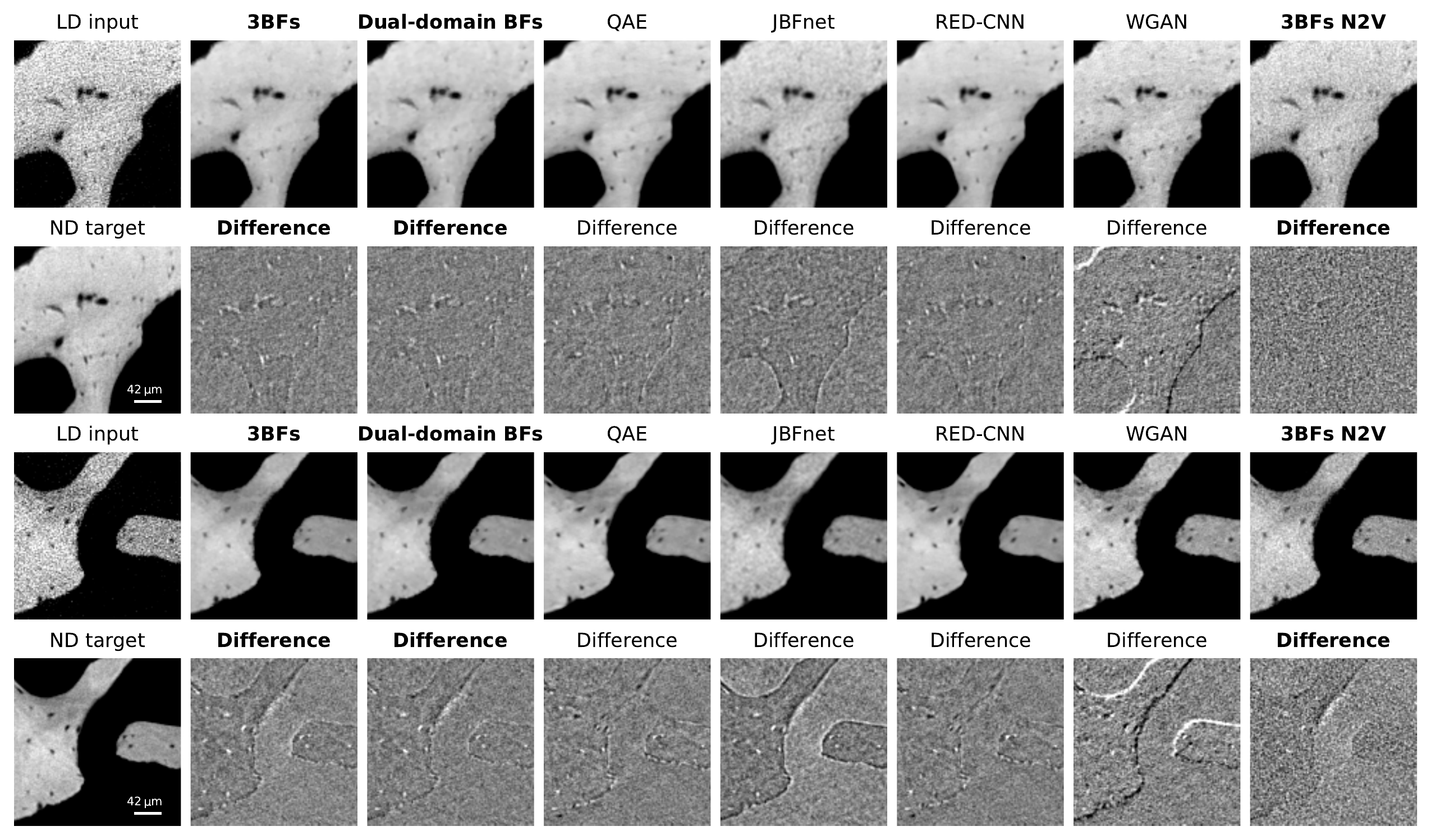}
    \caption{The first and third rows show denoising predictions from the XRM bone dataset in the ROIs illustrated in Fig. \ref{fig:Overview_ROIs} for all employed methods. Our proposed pipelines are highlighted in bold letters. The display window is $[0.25, 0.7]\,\text{arb.}\,\text{unit}$. The second and fourth rows present difference images between prediction and target where prediction artifacts like blurred edges become particularly visible. Difference images are plotted in the window $[-0.1, 0.1]\,\text{arb.}\,\text{unit}$.}
    \label{fig:results_rois_xrm}
    \vspace*{2ex}
\end{figure*}

We present qualitative denoising results on selected slices from both test datasets, visualized in Fig. \ref{fig:Overview_ROIs}. Multiple magnified regions of interest (ROIs) allow comparing the denoising performance on small image features of the Grand Challenge dataset for better visualization and are shown in Fig. \ref{fig:results_rois_mayo}. The additionally provided difference images (prediction - target) particularly highlight artifacts in the model predictions. Closer studying local features reveals over-smoothed results in low-contrast regions for the QAE and the RED-CNN compared to the normal-dose target, highlighted by orange circles and arrows. However, high-frequency details like edges are well preserved as there are few structures visible in the difference images. The predictions of bilateral filter-based pipelines, JBFnet, and WGAN visually appear more natural and closer to the target images, compared to the other approaches. However, blurred edges are visible in the difference images for JBFnet, WGAN, and the self-supervised bilateral filter pipeline (3BFs N2V). The predictions of the supervised-trained bilateral filter pipeline (3BFs) contain the noise pattern closest to the target ROI, while slightly stronger removing edges compared to QAE and RED-CNN. Quantitatively, RED-CNN, 3BFs, and QAE outperform the other models on the Grand Challenge dataset in terms of structural similarity index measure (SSIM) and peak signal-to-noise ratio (PSNR), presented in Table \ref{tab:quant_results_GC}.\\
The model predictions on the XRM bone dataset are depicted in Fig. \ref{fig:results_rois_xrm}, including the entire reconstruction pipeline, simultaneously denoising in sinogram and reconstruction domain (Dual-domain BFs). Visually, Dual-domain BFs, RED-CNN, and QAE outperform the other models in terms of noise removal and edge sensitivity. Predictions of the self-supervised bilateral filter pipeline preserve edges but still contain a substantial amount of noise. In contrast, the difference images of WGAN and JBFnet reveal more removed high-frequency details at edges, compared to all other methods. Quantitatively, denoising with three trained bilateral filters in the reconstruction domain performs slightly worse than QAE and RED-CNN in terms of SSIM and PSNR, as listed in Table \ref{tab:quant_results_xrm}. However, moving one of the bilateral filter layers into the sinogram domain improves the denoising performance to match the SSIM and PSNR of the best-performing deep models. All trainable bilateral filter-based pipelines use several orders of magnitude fewer parameters compared to the deep reference architectures. Simultaneously, quantitatively and qualitatively competitive denoising performance is achieved, outperforming multiple deep reference methods.

\subsection{Depth of the bilateral filter pipeline}

\begin{figure}[bth]
\centerline{\includegraphics[width=0.6\linewidth]{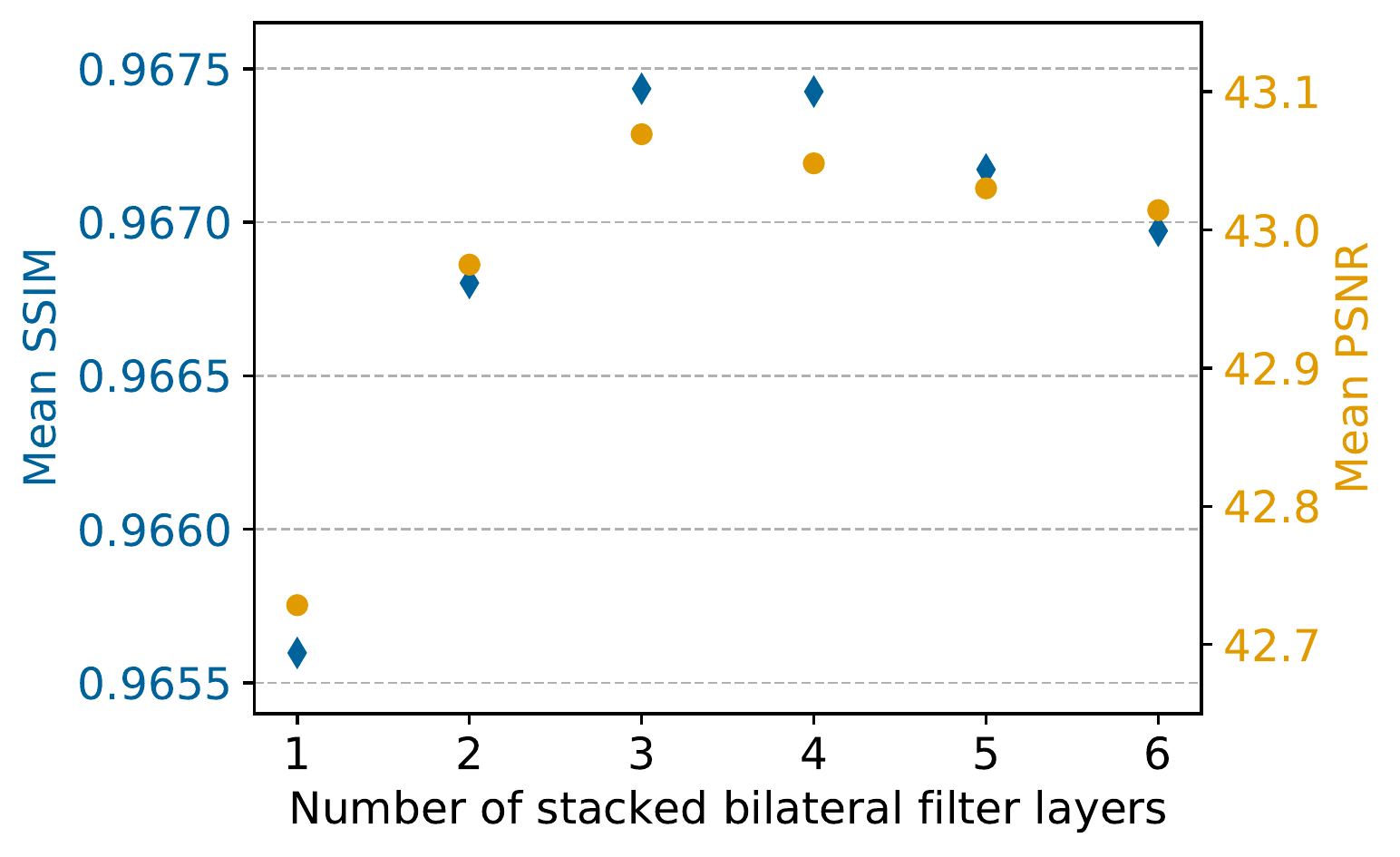}}
   \caption{The denoising performance of multiple stacked bilateral filter layers is compared based on the average SSIM and PSNR on the test patients of the 2016 Low Dose CT Grand Challenge dataset. Denoising with two or more trained layers is advantageous compared to only using a single layer.}
   \label{fig:N_Layer_Comparison}
   \vspace*{2ex}
\end{figure}

The benefit of iteratively removing noise with subsequent bilateral filters is investigated by training bilateral filter pipelines of different depths. Pipelines containing up to six filter layers are trained until convergence and tested on the Grand Challenge dataset. Mean SSIM and PSNR suggest that the denoising performance indeed benefits from stacking three filter layers, as shown in Fig. \ref{fig:N_Layer_Comparison}. Adding more filters slightly reduces the performance as more parameters have to properly converge to relatively small values to preserve edges.

\section{Discussion}
The proposed pipelines quantitatively and qualitatively show comparable denoising to state-of-the-art deep architectures. Although information from a single CT projection is spread through the whole volume, the appearance of noise in the reconstructed volume is a local phenomenon. Smart filtering within a finite neighborhood can, therefore, perform surprisingly well, as is the case for the bilateral filter. However, especially the choice of optimal range parameters $\sigma_r$ is very crucial for the denoising performance. If $\sigma_r$ is chosen too large, regions of constant attenuation are well restored, but edges are blurred concurrently, leading to degraded predictions. Optimizing the hyperparameters in a purely data-driven way can, therefore, leverage the applicability of the bilateral filter, as well as guarantee a near-optimal choice of parameters, as we demonstrated by achieving quantitative scores comparable to state-of-the-art methods.\\
We empirically found that the denoising performances of optimized pipelines are independent of the parameter initialization. For pipelines containing more than a single filter layer, the filter parameters first converge to very similar values between the layers until the loss is almost minimized. During further training, individual filter parameters start to converge to distinct values, slowly reducing the training loss further. During this fine-tuning phase, stacked bilateral filters learn to focus on different features, e.g., preserving edges or smoothing planes.\\
Compared to the ground truth targets, the predictions of RED-CNN and QAE visually appear over-smoothed in low-contrast regions, which can remove features being beneficial for the physician. Whereas supervised training of the bilateral filter-based pipelines performs comparable to deep reference models on both investigated datasets, the self-supervised approach using the Noise2Void technique does not fully converge to the optimal parameter set. In general, improved image quality can be recognized, however, the presented difference images identify blurred edges due to insufficiently learned filter parameters. If noise is not fully distinguishable from content, the range parameters $\sigma_r$ do not properly converge to the optimum values. In their work, Krull \textit{et al.} \cite{krull2019noise2void} already identified this limitation of their Noise2Void algorithm for images with inherently high irregularities.\\

\begin{figure}[tbh]
\centerline{\includegraphics[width=0.7\linewidth]{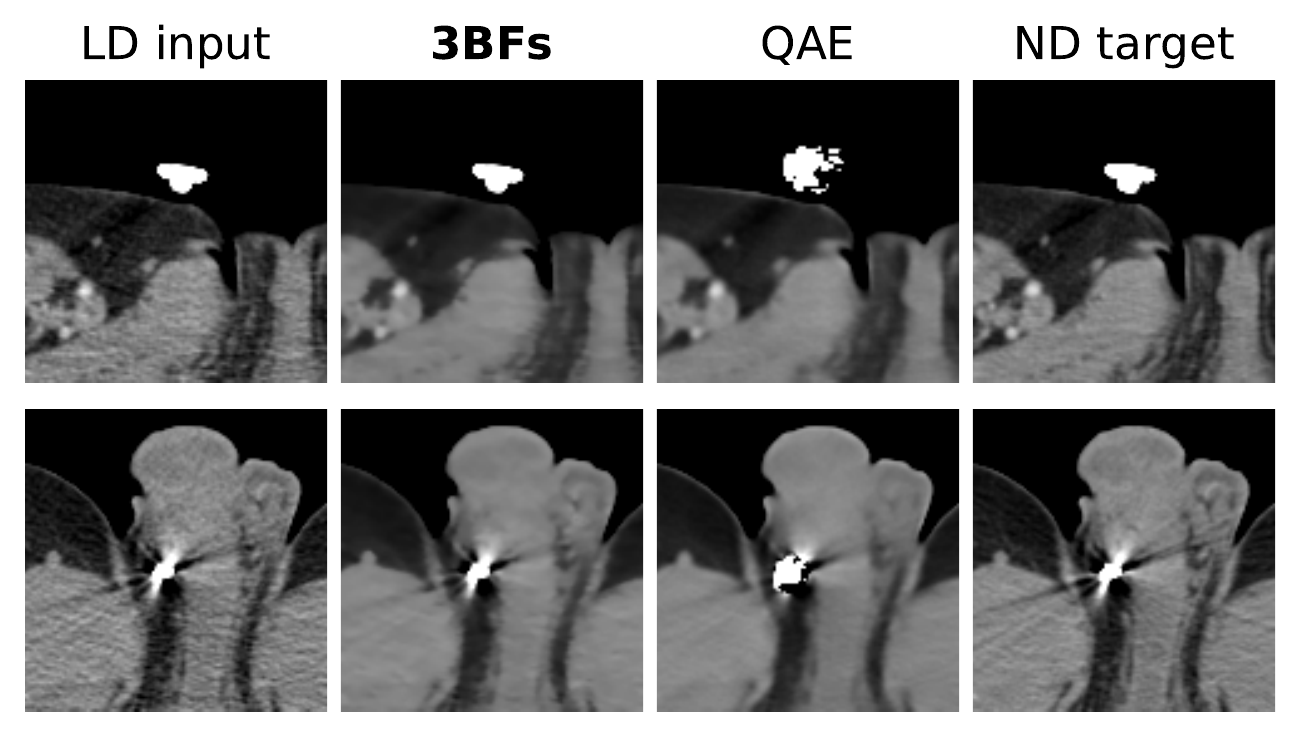}}
   \caption{Example of failing network predictions of the trained deep QAE model on strongly absorbing objects in the scan of patient \textit{L006} and \textit{L193} from the \textit{Low Dose CT Image and Projection Dataset} \cite{moen2021low}. The display window is $[-150, 250]\,\text{HU}$.}
   \label{fig:Results_ROI_Stability}
   \vspace*{2ex}
\end{figure}

Approximation and estimation error are commonly combined to estimate the risk bound of neural networks. The two contributions refer first to the distance between target function and closest neural network function and second to the distance between the estimated and the ideal network function \cite{barron1994approximation}. Our quantitative and visual results show that deep neural networks and trainable bilateral filter pipelines can both well predict denoised reconstructions after extensive training and, therefore, show comparable approximation risk. However, we think trainable bilateral filter layers have a considerably smaller estimation error due to the restricted influence of the few trained parameters. The decreased error bound can be explained by treating the filter algorithm similar to a known operator \cite{maier2019, maier2018precision}. In contrast, deep neural network-based approaches can learn network functions that only work well within the training data distribution but fail to produce reliable results for other samples from the CT data distribution, which results in a larger estimation error.\\
In a clinical environment, data integrity at any point in the medical data processing pipeline is crucial for a reliable diagnosis from the physician. During training, deep models are optimized on a finite number of training samples to extract features generalizing over the entire distribution of data. However, in a clinical environment, data can be acquired which is not properly represented by the finite training data distribution. Image processing algorithms must be able to handle such samples to avoid failing downstream tasks, including the diagnosis by the physician. Fig. \ref{fig:Results_ROI_Stability} exhibits failing network predictions of one deep reference method on out-of-distribution CT slices around medical implants in abdomen scans of the \textit{Low Dose CT Image and Projection Dataset} \cite{moen2021low}. The reference method was before trained on the 2016 Low Dose CT Grand Challenge training dataset until convergence and achieves state-of-the-art performance on the test data, as presented in Table \ref{tab:quant_results_GC}. A different predicted artifact of a deep reference method is highlighted in Fig. \ref{fig:results_rois_mayo} with red arrows. Here, a shadow-like structure right next to the bone region is introduced which is also visible in the difference image. By algorithmic design, trainable bilateral filter layers only filter in a finite neighborhood with well-defined Gaussian kernels and can not produce such artifacts in their predictions. The inherent lack of complex feature abstraction in the proposed filter layer enforces proximity between prediction and measured ground truth and is advantageous for maintaining reliable data compared to the nontransparent data processing performed in deep state-of-the-art denoising models. Additionally, the low amount of required training data makes trainable bilateral filter layers convenient to use and easy to integrate into complex data processing pipelines.

\section{Conclusions}
We present a trainable, fully differentiable bilateral filter layer that can be directly incorporated in any deep architecture using the \textit{PyTorch} framework with GPU acceleration. We show that trainable bilateral filter layers with only four parameters can achieve state-of-the-art performance of deep neural networks with hundred of thousands of trainable parameters on the low-dose CT denoising task, while performing fully comprehensible data processing and producing physically reliable predictions at any time. The low number of parameters allows training with little data, self-supervised training using the Noise2Void scheme, and incorporating the filter layer at multiple locations in a CT reconstruction pipeline to further leverage denoising performance. In summary, trainable bilateral filters allow extensive dose reduction while maintaining high image quality without sacrificing data integrity.

\section{Acknowledgments}
The research leading to these results has received funding from the European Research Council (ERC) under the European Union’s Horizon 2020 research and innovation program (ERC Grant No. 810316).

\section{Conflict of Interest}
The authors have no relevant conflicts of interest to disclose.

% following only if there is an appendix
\section*{Appendix I: Analytical derivative of the bilateral filter}
\addcontentsline{toc}{section}{\numberline{}Appendix I}
\label{sec:Appendix_derivative}

We aim to make the spatial and range hyper parameters $\sigma_s$ and $\sigma_r$ of the bilateral filter trainable in a gradient descent algorithm. In order to update the parameters during optimization the derivative of the loss function $L$ with respect to each parameter $\sigma_\gamma$ is required
\begin{align}
    \frac{\partial L}{\partial \sigma_\gamma} = \frac{\partial L}{\partial \mathbf{\hat{Y}}} \frac{\partial \mathbf{\hat{Y}}}{\partial \sigma_\gamma}  = \sum_k \frac{\partial L}{\partial \hat{Y}_k} \frac{\partial \hat{Y}_k}{\partial \sigma_\gamma}.
\end{align}
Following the definition of the bilateral filter in (\ref{eq:DefBF}), the derivative $\frac{\partial \hat{Y}_k}{\partial \sigma_\gamma}$ yields
\begin{align}
    \frac{\partial \hat{Y}_k}{\partial \sigma_\gamma} = - w_k^{-2} \alpha_k \frac{\partial w_k}{\partial \sigma_\gamma} + w_k^{-1} \frac{\partial \alpha_k}{\partial \sigma_\gamma}
    \label{eq:DerivativeSigI}
\end{align}
with
\begin{align}
    \frac{\partial w_k}{\partial \sigma_\gamma} &= \sum_{n \in \mathcal{N}} \frac{\partial}{\partial \sigma_\gamma} G_{\sigma_s}(\mathbf{p_k} - \mathbf{p_n}) G_{\sigma_r}(X_k - X_n),\\
    \frac{\partial \alpha_k}{\partial \sigma_\gamma} &= \sum_{n \in \mathcal{N}} X_n \frac{\partial}{\partial \sigma_\gamma} G_{\sigma_s}(\mathbf{p_k} - \mathbf{p_n}) G_{\sigma_r}(X_k - X_n),
\end{align}
and the derivative of the Gaussian kernel
\begin{align}
    \frac{\partial}{\partial \sigma_\gamma} G_{\sigma}(c) = G_{\sigma}(c) \frac{c^2}{\sigma_\gamma^3}.
\end{align}
Additionally, the gradient with respect to every voxel of the noisy input volume $X_i$ needs to be derived, as it should be possible to propagate a loss into previous filter layers during backpropagation
\begin{align}
    \frac{\partial L}{\partial X_i} = \frac{\partial L}{\partial \mathbf{\hat{Y}}} \frac{\partial \mathbf{\hat{Y}}}{\partial X_i} = \sum_k \frac{\partial L}{\partial \hat{Y}_k} \frac{\partial \hat{Y}_k}{\partial X_i}.
    \label{eq:BackpropConvBF}
\end{align}
The derivative of the output voxel $\hat{Y}_k$ with respect to the input voxel $X_i$ yields
\begin{align}
    \frac{\partial \hat{Y}_k}{\partial X_i} = - w_k^{-2} \alpha_k \frac{\partial w_k}{\partial X_i} + w_k^{-1} \frac{\partial \alpha_k}{\partial X_i}
\end{align}
by again using the definition of the bilateral filter from (\ref{eq:DefBF}) and applying the product rule analog to (\ref{eq:DerivativeSigI}).\\
The filtered output voxel $\hat{Y}_k$ is dependent on both input voxels $X_k$ and $X_n$. To carry out the derivative with respect to $X_i$ the two cases $k \neq i$ and $k=i$ are distinguished.\\
\\
\textbf{Case 1: (\boldmath$k \neq i$)}
\begin{align}
    \begin{split}
        \left.\frac{\partial w_k}{\partial X_i}\right\vert_{k \neq i} &= \,\sum_{n \in \mathcal{N}} G_{\sigma_s}(\mathbf{p_k} - \mathbf{p_n}) \frac{\partial}{\partial X_i} G_{\sigma_r}(X_k - X_n)\\
        &= \,G_{\sigma_s}(\mathbf{p_k} - \mathbf{p_i}) G_{\sigma_r}(X_k - X_i) \frac{X_k - X_i}{\sigma_r^2}
        \label{eq:K_not_I_1}
    \end{split}
\end{align}
\begin{align}
    \begin{split}
        \left.\frac{\partial \alpha_k}{\partial X_i}\right\vert_{k \neq i} =& \,\sum_{n \in \mathcal{N}} G_{\sigma_s}(\mathbf{p_k} - \mathbf{p_n}) \frac{\partial}{\partial X_i} G_{\sigma_r}(X_k - X_n) X_n\\
        =& \,G_{\sigma_s}(\mathbf{p_k} - \mathbf{p_i}) \left[ \left(\frac{\partial}{\partial X_i} G_{\sigma_r}(X_k - X_i)\right) X_i + G_{\sigma_r}(X_k - X_i) \left(\frac{\partial}{\partial X_i} X_i\right) \right]\\
        =& \,G_{\sigma_s}(\mathbf{p_k} - \mathbf{p_i}) \cdot G_{\sigma_r}(X_k - X_i) \left[ \frac{X_k - X_i}{\sigma_r^2} X_i + 1 \right]
        \label{eq:K_not_I_2}
    \end{split}
\end{align}
In both expressions only the $i$th term of the sum ($n=i$) contributes.\\
\\
\textbf{Case 2: (\boldmath$k = i$)}
\begin{align}
    \begin{split}
        \left.\frac{\partial w_k}{\partial X_i}\right\vert_{k = i} =& \,\frac{\partial}{\partial X_i} \sum_{n \in \mathcal{N}} G_{\sigma_s}(\mathbf{p_i} - \mathbf{p_n}) G_{\sigma_r}(X_i - X_n)\\
        =& \,\frac{\partial}{\partial X_i} \left[ \underbrace{G_{\sigma_s}(\mathbf{p_i} - \mathbf{p_i}) G_{\sigma_r}(X_i - X_i)}_{\substack{=1}} + \sum_{n \in \mathcal{N}, n \neq i} G_{\sigma_s}(\mathbf{p_i} - \mathbf{p_n}) G_{\sigma_r}(X_i - X_n) \right]\\
        =& \,\sum_{n \in \mathcal{N}, n \neq i} G_{\sigma_s}(\mathbf{p_i} - \mathbf{p_n}) \frac{\partial}{\partial X_i} G_{\sigma_r}(X_i - X_n)\\
        =& \,\sum_{n \in \mathcal{N}} G_{\sigma_s}(\mathbf{p_i} - \mathbf{p_n}) G_{\sigma_r}(X_i - X_n) \frac{X_n - X_i}{\sigma_r^2}
        \label{eq:K_is_I_1}
    \end{split}
\end{align}
\begin{align}
    \begin{split}
        \left.\frac{\partial \alpha_k}{\partial X_i}\right\vert_{k = i} =& \,\frac{\partial}{\partial X_i} \sum_{n \in \mathcal{N}} G_{\sigma_s}(\mathbf{p_i} - \mathbf{p_n}) G_{\sigma_r}(X_i - X_n) X_n\\
        =& \,\frac{\partial}{\partial X_i} \left[ \underbrace{G_{\sigma_s}(\mathbf{p_i} - \mathbf{p_i}) G_{\sigma_r}(X_i - X_i)}_{\substack{=1}} X_i + \sum_{n \in \mathcal{N}, n \neq i} G_{\sigma_s}(\mathbf{p_i} - \mathbf{p_n}) G_{\sigma_r}(X_i - X_n) X_n \right]\\
        =& \,1 + \sum_{n \in \mathcal{N}, n \neq i} \left[ G_{\sigma_s}(\mathbf{p_i} - \mathbf{p_n}) \cdot G_{\sigma_r}(X_i - X_n) X_n \frac{X_n - X_i}{\sigma_r^2}\right]\\
        =& \,1 + \sum_{n \in \mathcal{N}} \left[ G_{\sigma_s}(\mathbf{p_i} - \mathbf{p_n}) \cdot G_{\sigma_r}(X_i - X_n) X_n \frac{X_n - X_i}{\sigma_r^2} \right]
        \label{eq:K_is_I_2}
    \end{split}
\end{align}
Note that in the last steps the sum can be carried out over the entire neighborhood $N$ as the contribution of the term $n=i$ is zero respectively.

\section*{Appendix II: Implementation}
\addcontentsline{toc}{section}{\numberline{}Appendix II}
\label{sec:Appendix_implementation}

In practice, (\ref{eq:BackpropConvBF}) describes a convolution of the kernel $\frac{\partial \hat{Y}_k}{\partial X_i}$ with the backpropagated loss $\frac{\partial L}{\partial \hat{Y}_k}$ in a neighborhood denoted by $k$ which is given by the finite kernel size---analog to a conventional convolutional layer. However, due to the dependency of the kernel on two voxels of the input volume $X_k$ and $X_i$, the analytical derivative of the convolutional kernel is more elaborate. When calculating the sum in (\ref{eq:BackpropConvBF}), both (\ref{eq:K_not_I_1}) and (\ref{eq:K_not_I_2}) are used for the contributions $k \neq i$ and only the term $k = i$ is derived in (\ref{eq:K_is_I_1}) and (\ref{eq:K_is_I_2}). Note that for faster computation the sums in (\ref{eq:K_is_I_1}) and (\ref{eq:K_is_I_2}) can be pre-calculated in the forward pass of the filtering. The finite spatial kernel is calculated by taking into account voxels covering five times the spatial standard deviation in each spatial dimension, but at least $5 \times 5 \times 5$ voxels. This could affect the performance for large spatial kernels with $\sigma_s \gg 1$, but was never a relevant factor in any of our experiments.\\
To validate the correct implementation of the filter derivative, we compared the analytical gradient, provided by the backward pass of the filter layer, with a numerical approximation of the gradient, computed via small finite differences. Our public repository contains a gradient check script using the \textit{torch.autograd.gradcheck} function from the \textit{PyTorch} framework.

% following only if there is an appendix
%\section*{Appendix}
%\addcontentsline{toc}{section}{\numberline{}Appendix}
%Appendix text goes here if needed.

\section*{References}
\addcontentsline{toc}{section}{\numberline{}References}
\vspace*{-20mm}

% Following assumes you are using bibtex. However, for submission to the
% journal you MUST explicitly INCLUDE THE REFERENCES IN THE TEX FILE. 
% In that case you need the following

% \begin{thebibliography}{10}
% insert the .bbl file generated by bibtex here
	%This will be a series of entries from your .bib file formatted
	%something like
	%\bibitem{Me09}
        %{I.~Meijsing, B.~W.~Raaymakers, A.~J.~E.~Raaijmakers \it et al.},
        %\newblock {Dosimetry for the MRI accelerator: the impact of a 
	%magnetic field on the response of a Farmer NE2571 ionization chamber},
        %\newblock Phys. Med. Biol. {\bf 54}, 2993 -- 3002 (2009).

% \end{thebibliography}

% The following is when using bibtex and picks up the example.bib file

%\bibliography{Explicit address of .bib file}
\bibliography{example}      %example.bib is on the same directory

\begin{thebibliography}{10}

\bibitem{boone2012radiation}
J.~M. Boone, W.~R. Hendee, M.~F. McNitt-Gray, and S.~E. Seltzer,
\newblock Radiation exposure from CT scans: how to close our knowledge gaps,
  monitor and safeguard exposure—proceedings and recommendations of the
  Radiation Dose Summit, sponsored by NIBIB, February 24--25, 2011,
\newblock Radiology {\bf 265}, 544--554 (2012).

\bibitem{barrett1976statistical}
H.~H. Barrett, S.~Gordon, and R.~Hershel,
\newblock Statistical limitations in transaxial tomography,
\newblock Computers in biology and medicine {\bf 6}, 307--323 (1976).

\bibitem{maier2015gpu}
A.~Maier and R.~Fahrig,
\newblock GPU denoising for computed tomography,
\newblock Graphics Processing Unit-Based High Performance Computing in
  Radiation Therapy {\bf 113} (2015).

\bibitem{kelm2009optimizing}
Z.~S. Kelm, D.~Blezek, B.~Bartholmai, and B.~J. Erickson,
\newblock Optimizing non-local means for denoising low dose CT,
\newblock in {\em 2009 IEEE International Symposium on Biomedical Imaging: From
  Nano to Macro}, pages 662--665, IEEE, 2009.

\bibitem{maier2011three}
A.~Maier, L.~Wigstr{\"o}m, H.~G. Hofmann, J.~Hornegger, L.~Zhu, N.~Strobel, and
  R.~Fahrig,
\newblock Three-dimensional anisotropic adaptive filtering of projection data
  for noise reduction in cone beam CT,
\newblock Med. Phys. {\bf 38}, 5896--5909 (2011).

\bibitem{manduca2009projection}
A.~Manduca, L.~Yu, J.~D. Trzasko, N.~Khaylova, J.~M. Kofler, C.~M. McCollough,
  and J.~G. Fletcher,
\newblock Projection space denoising with bilateral filtering and CT noise
  modeling for dose reduction in CT,
\newblock Med. Phys. {\bf 36}, 4911--4919 (2009).

\bibitem{manhart2014guided}
M.~Manhart, R.~Fahrig, J.~Hornegger, A.~Doerfler, and A.~Maier,
\newblock Guided noise reduction for spectral CT with energy-selective photon
  counting detectors,
\newblock in {\em Proceedings of the Third CT Meeting}, pages 91--94, 2014.

\bibitem{han2010algorithm}
X.~Han, J.~Bian, D.~R. Eaker, T.~L. Kline, E.~Y. Sidky, E.~L. Ritman, and
  X.~Pan,
\newblock Algorithm-enabled low-dose micro-CT imaging,
\newblock IEEE Transactions on Medical Imaging {\bf 30}, 606--620 (2010).

\bibitem{beister2012iterative}
M.~Beister, D.~Kolditz, and W.~A. Kalender,
\newblock Iterative reconstruction methods in X-ray CT,
\newblock Physica Medica {\bf 28}, 94--108 (2012).

\bibitem{gilbert1972iterative}
P.~Gilbert,
\newblock Iterative methods for the three-dimensional reconstruction of an
  object from projections,
\newblock Journal of theoretical biology {\bf 36}, 105--117 (1972).

\bibitem{chen2017low}
H.~Chen, Y.~Zhang, M.~K. Kalra, F.~Lin, Y.~Chen, P.~Liao, J.~Zhou, and G.~Wang,
\newblock Low-Dose CT With a Residual Encoder-Decoder Convolutional Neural
  Network,
\newblock IEEE Transactions on Medical Imaging {\bf 36}, 2524--2535 (2017).

\bibitem{patwari2020}
M.~Patwari, R.~Gutjahr, R.~Raupach, and A.~Maier,
\newblock JBFnet - Low Dose CT Denoising by Trainable Joint Bilateral
  Filtering,
\newblock in {\em International Conference on Medical Image Computing and
  Computer-Assisted Intervention---MICCAI 2020}, pages 506--515, Springer,
  2020.

\bibitem{fan2019quadratic}
F.~Fan, H.~Shan, M.~K. Kalra, R.~Singh, G.~Qian, M.~Getzin, Y.~Teng, J.~Hahn,
  and G.~Wang,
\newblock Quadratic autoencoder (Q-AE) for low-dose CT denoising,
\newblock IEEE Transactions on Medical Imaging {\bf 39}, 2035--2050 (2019).

\bibitem{yang2018low}
Q.~Yang, P.~Yan, Y.~Zhang, H.~Yu, Y.~Shi, X.~Mou, M.~K. Kalra, Y.~Zhang,
  L.~Sun, and G.~Wang,
\newblock Low-Dose CT Image Denoising Using a Generative Adversarial Network
  With Wasserstein Distance and Perceptual Loss,
\newblock IEEE Transactions on Medical Imaging {\bf 37}, 1348--1357 (2018).

\bibitem{kang2018deep}
E.~Kang, W.~Chang, J.~Yoo, and J.~C. Ye,
\newblock Deep convolutional framelet denosing for low-dose CT via wavelet
  residual network,
\newblock IEEE Transactions on Medical Imaging {\bf 37}, 1358--1369 (2018).

\bibitem{kang2017deep}
E.~Kang, J.~Min, and J.~C. Ye,
\newblock A deep convolutional neural network using directional wavelets for
  low-dose X-ray CT reconstruction,
\newblock Med. Phys. {\bf 44}, e360--e375 (2017).

\bibitem{ketcha2021}
M.~D. Ketcha, M.~Marrama, A.~Souza, A.~Uneri, P.~Wu, X.~Zhang, P.~A. Helm, and
  J.~H. Siewerdsen,
\newblock {Sinogram + image domain neural network approach for metal artifact
  reduction in low-dose cone-beam computed tomography},
\newblock Journal of Medical Imaging {\bf 8}, 1--16 (2021).

\bibitem{yuan2019adversarial}
X.~Yuan, P.~He, Q.~Zhu, and X.~Li,
\newblock Adversarial Examples: Attacks and Defenses for Deep Learning,
\newblock IEEE Transactions on Neural Networks and Learning Systems {\bf 30},
  2805--2824 (2019).

\bibitem{antun2020instabilities}
V.~Antun, F.~Renna, C.~Poon, B.~Adcock, and A.~C. Hansen,
\newblock On instabilities of deep learning in image reconstruction and the
  potential costs of AI,
\newblock Proceedings of the National Academy of Sciences {\bf 117},
  30088--30095 (2020).

\bibitem{huang2018some}
Y.~Huang, T.~W{\"u}rfl, K.~Breininger, L.~Liu, G.~Lauritsch, and A.~Maier,
\newblock Some investigations on robustness of deep learning in limited angle
  tomography,
\newblock in {\em International Conference on Medical Image Computing and
  Computer-Assisted Intervention---MICCAI 2018}, pages 145--153, Springer,
  2018.

\bibitem{tomasi1998bilateral}
C.~Tomasi and R.~Manduchi,
\newblock Bilateral filtering for gray and color images,
\newblock in {\em Sixth international conference on computer vision}, pages
  839--846, IEEE, 1998.

\bibitem{chen2013optimization}
Y.~Chen and Y.~Shu,
\newblock Optimization of bilateral filter parameters via chi-square unbiased
  risk estimate,
\newblock IEEE Signal Processing Letters {\bf 21}, 97--100 (2013).

\bibitem{anoop2019medical}
V.~Anoop and P.~R. Bipin,
\newblock Medical image enhancement by a bilateral filter using optimization
  technique,
\newblock Journal of medical systems {\bf 43}, 1--12 (2019).

\bibitem{kishan2012optimal}
H.~Kishan and C.~S. Seelamantula,
\newblock Optimal parameter selection for bilateral filters using Poisson
  Unbiased Risk Estimate,
\newblock in {\em 2012 19th IEEE International Conference on Image Processing},
  pages 121--124, IEEE, 2012.

\bibitem{dai2018content}
T.~Dai, Y.~Zhang, L.~Dong, L.~Li, X.~Liu, and S.~Xia,
\newblock Content-Aware Bilateral Filtering,
\newblock in {\em 2018 IEEE Fourth International Conference on Multimedia Big
  Data (BigMM)}, pages 1--6, IEEE, 2018.

\bibitem{peng2010bilateral}
H.~Peng and R.~Rao,
\newblock Bilateral kernel parameter optimization by risk minimization,
\newblock in {\em 2010 IEEE International Conference on Image Processing},
  pages 3293--3296, IEEE, 2010.

\bibitem{jampani2016learning}
V.~Jampani, M.~Kiefel, and P.~V. Gehler,
\newblock Learning sparse high dimensional filters: Image filtering, dense crfs
  and bilateral neural networks,
\newblock in {\em Proceedings of the IEEE Conference on Computer Vision and
  Pattern Recognition}, pages 4452--4461, 2016.

\bibitem{kang2021low}
W.~Kang and M.~Patwari,
\newblock Low Dose Helical CBCT denoising by using domain filtering with deep
  reinforcement learning,
\newblock arXiv preprint arXiv:2104.00889  (2021).

\bibitem{patwari2020low}
M.~Patwari, R.~Gutjahr, R.~Raupach, and A.~Maier,
\newblock Low Dose CT Denoising via Joint Bilateral Filtering and Intelligent
  Parameter Optimization,
\newblock arXiv preprint arXiv:2007.04768  (2020).

\bibitem{pytorch2019}
A.~Paszke et~al.,
\newblock PyTorch: An Imperative Style, High-Performance Deep Learning Library,
\newblock in {\em Proc. NeurIPS}, pages 8024--8035, Curran Associates, Inc.,
  2019.

\bibitem{syben2019}
C.~Syben, M.~Michen, B.~Stimpel, S.~Seitz, S.~Ploner, and A.~K. Maier,
\newblock PYRO-NN: Python reconstruction operators in neural networks,
\newblock Med. Phys. {\bf 46}, 5110--5115 (2019).

\bibitem{krull2019noise2void}
A.~Krull, T.-O. Buchholz, and F.~Jug,
\newblock Noise2void-learning denoising from single noisy images,
\newblock in {\em Proceedings of the IEEE/CVF Conference on Computer Vision and
  Pattern Recognition}, pages 2129--2137, 2019.

\bibitem{maier2019}
A.~Maier, C.~Syben, B.~Stimpel, T.~W{\"u}rfl, M.~Hoffmann, F.~Schebesch, W.~Fu,
  L.~Mill, L.~Kling, and S.~Christiansen,
\newblock Learning with known operators reduces maximum error bounds,
\newblock Nature machine intelligence {\bf 1}, 373--380 (2019).

\bibitem{maier2018precision}
A.~Maier, F.~Schebesch, C.~Syben, T.~W{\"u}rfl, S.~Steidl, J.-H. Choi, and
  R.~Fahrig,
\newblock Precision learning: towards use of known operators in neural
  networks,
\newblock in {\em 2018 24th International Conference on Pattern Recognition
  (ICPR)}, pages 183--188, IEEE, 2018.

\bibitem{mccollough2017low}
C.~H. McCollough et~al.,
\newblock Low-dose CT for the detection and classification of metastatic liver
  lesions: results of the 2016 low dose CT grand challenge,
\newblock Med. Phys. {\bf 44}, e339--e352 (2017).

\bibitem{pybind11}
W.~Jakob,
\newblock pybind11 -- Seamless operability between C++11 and Python, 2021.

\bibitem{thies2021differentiable}
M.~Thies,
\newblock Differentiable reconstruction for x-ray microscopy data,
\newblock Code Ocean  (2021).

\bibitem{feldkamp1984practical}
L.~A. Feldkamp, L.~C. Davis, and J.~W. Kress,
\newblock Practical cone-beam algorithm,
\newblock Josa a {\bf 1}, 612--619 (1984).

\bibitem{hannah2010}
K.~M. Hannah, C.~D. Thomas, J.~G. Clement, F.~{De Carlo}, and A.~G. Peele,
\newblock {Bimodal distribution of osteocyte lacunar size in the human femoral
  cortex as revealed by micro-CT},
\newblock Bone {\bf 47}, 866--871 (2010).

\bibitem{gruneboom20191}
A.~Gr{\"{u}}neboom et~al.,
\newblock A network of trans-cortical capillaries as mainstay for blood
  circulation in long bones,
\newblock Nature Metabolism {\bf 1}, 236--250 (2019).

\bibitem{gruneboom20192}
A.~Gr{\"{u}}neboom, L.~Kling, S.~Christiansen, L.~Mill, A.~Maier, K.~Engelke,
  H.~H. Quick, G.~Schett, and M.~Gunzer,
\newblock {Next-generation imaging of the skeletal system and its blood
  supply},
\newblock Nature Reviews Rheumatology {\bf 15}, 533--549 (2019).

\bibitem{gruber2008introduction}
R.~Gruber, P.~Pietschmann, and M.~Peterlik,
\newblock Introduction to bone development, remodelling and repair,
\newblock in {\em Radiology of osteoporosis}, pages 1--23, Springer, 2008.

\bibitem{buenzli2015}
P.~R. Buenzli and N.~A. Sims,
\newblock {Quantifying the osteocyte network in the human skeleton},
\newblock Bone {\bf 75}, 144--150 (2015).

\bibitem{mill2019towards}
L.~Mill, L.~Kling, A.~Gr{\"u}neboom, G.~Schett, S.~Christiansen, and A.~Maier,
\newblock Towards In-Vivo X-Ray Nanoscopy,
\newblock in {\em Bildverarbeitung f{\"u}r die Medizin 2019}, pages 251--256,
  Springer, 2019.

\bibitem{huang2021semi}
Y.~Huang, L.~Mill, R.~Stoll, L.~Kling, O.~Aust, F.~Wagner, A.~Gr{\"u}neboom,
  G.~Schett, S.~Christiansen, and A.~Maier,
\newblock Semi-permeable Filters for Interior Region of Interest Dose Reduction
  in X-ray Microscopy.,
\newblock in {\em Bildverarbeitung f{\"u}r die Medizin}, pages 61--66, 2021.

\bibitem{yu2012development}
L.~Yu, M.~Shiung, D.~Jondal, and C.~H. McCollough,
\newblock Development and validation of a practical lower-dose-simulation tool
  for optimizing computed tomography scan protocols,
\newblock Journal of computer assisted tomography {\bf 36}, 477--487 (2012).

\bibitem{moen2021low}
T.~R. Moen, B.~Chen, D.~R. Holmes~III, X.~Duan, Z.~Yu, L.~Yu, S.~Leng, J.~G.
  Fletcher, and C.~H. McCollough,
\newblock Low-dose CT image and projection dataset,
\newblock Med. Phys. {\bf 48}, 902--911 (2021).

\bibitem{barron1994approximation}
A.~R. Barron,
\newblock Approximation and estimation bounds for artificial neural networks,
\newblock Machine learning {\bf 14}, 115--133 (1994).

\end{thebibliography}
% above points to where we find the master reference list
% and also causes the bibliography to be printed

% When creating your bibliography you should run bibtex on your local
% computer after running pdflatex on your .tex file. bibtex will
% generate a .bbl file.
% Copy the contents of this .bbl file into your main latex document,
% replacing the "\bibliography" command which was pointing at your .bib file.

% following defines style of .bbl file 

%\bibliographystyle{explicit relative path to medphy.bst}
\bibliographystyle{medphy.bst}    %if this is installed on your system,
				    %it is not essential to have the    ./

% Note that you need to typeset once, then run bibtex, then typeset another
% two times to get the references working properly.

\end{document}